# Universal conductance fluctuations and phase-coherent transport in a semiconductor $Bi_2O_2Se$ nanoplate with strong spin-orbit interaction

Mengmeng Meng,[1] Shaoyun Huang,[1,*] Congwei Tan,[2,3] Jinxiong Wu,[2] Xiaobo Li,[1,3] Hailin Peng[2,*] and H. Q. Xu[1,4,5,*]

[1]*Beijing Key Laboratory of Quantum Devices, Key Laboratory for the Physics and Chemistry of Nanodevices and Department of Electronics, Peking University, Beijing 100871, China*

[2]*Center for Nanochemistry, Beijing National Laboratory for Molecular Sciences (BNLMS), College of Chemistry and Molecular Engineering, Peking University, Beijing 100871, China*

[3]*Academy for Advanced Interdisciplinary Studies, Peking University, Beijing 100871, China*

[4]*Beijing Academy of Quantum Information Sciences, West Bld. #3, No.10 Xibeiwang East Rd., Haidian District, Beijing 100193, China*

[5]*NanoLund and Division of Solid State Physics, Lund University, Box 118, S-221 00, Lund, Sweden*

We report on phase-coherent transport studies of a $Bi_2O_2Se$ nanoplate and on observation of universal conductance fluctuations and spin-orbit interaction induced reduction in fluctuation amplitude in the nanoplate. Thin-layered $Bi_2O_2Se$ nanoplates are grown by chemical vapor deposition (CVD) and transport measurements are made on a Hall-bar device fabricated from a CVD-grown nanoplate. The measurements show weak antilocalization at low magnetic fields at low temperatures, as a result of spin-orbit interaction, and a crossover toward weak localization with increasing temperature. Temperature dependences of characteristic transport lengths, such as spin relaxation length, phase coherence length, and mean free path, are extracted from the low-field measurement data. Universal conductance fluctuations are visible in the low-temperature magnetoconductance over a large range of magnetic fields and the phase coherence length extracted from the autocorrelation function is in consistence with the result obtained from the weak localization analysis. More importantly, we find a strong reduction in amplitude of the universal conductance fluctuations and show that the results agree with the analysis assuming strong spin-orbit interaction in the $Bi_2O_2Se$ nanoplate.

*Corresponding authors: Professor H. Q. Xu (hqxu@pku.edu.cn), Dr. Shaoyun Huang (syhuang@pku.edu.cn), and Professor Hailin Peng (hlpeng@pku.edu.cn)

Since the discovery of graphene,[1, 2] extensive research has been focused on realizations of various two-dimensional (2D) material systems and on their potential applications in electronics, optoelectronics, and nanoelectromechanical systems.[3, 4] Recently, a new type of 2D material, semiconductor $Bi_2O_2Se$ nanoplate, has been successfully synthesized via chemical vapor deposition (CVD).[5-7] These CVD-grown $Bi_2O_2Se$ nanoplates can be stored in ambient environment for a considerably long period of time without crystal quality degrading.[7, 8] $Bi_2O_2Se$ has an indirect bandgap of ~0.8 eV and field-effect devices made from non-encapsulated $Bi_2O_2Se$ nanoplates exhibit high electron mobility of ~450 $cm^2\ V^{-1}\ s^{-1}$ at room temperature.[7] Based on their suitable bandgap and high carrier mobility, $Bi_2O_2Se$ nanoplates have been used to construct infrared photodetectors with fast responsivity and high sensitivity.[9-11] $Bi_2O_2Se$ also exhibits excellent low-temperature transport properties. For example, high electron mobility at low-temperature has been reported in $Bi_2O_2Se$ layers (~$2.9\times10^4$ to ~$2.8\times10^5$ $cm^2\ V^{-1}s^{-1}$) and, as a result, Shubnikov-de Haas (SdH) oscillations have been detected in these materials.[7, 8, 12] Recently, we demonstrated the existence of strong spin-orbit interaction in $Bi_2O_2Se$ nanoplates and observed weak antilocalization (WAL), a signature of spin-orbit interaction, and its crossover to weak localization (WL) with increasing temperature or decreasing gate voltage.[13] Tunable WAL and strong spin-orbital interaction have also been observed in other 2D material systems, such as monolayer and a few-layer $MoS_2$,[14, 15] $WSe_2$ nanoflakes,[16] GaSe thin films,[17] InSe nanoflakes.[18, 19] Strong spin-orbital interaction characteristics make these 2D materials very interesting for applications in quantum electronics and spintronics, and for exploring new physics. However, as to our best knowledge, no report has been made on observations of universal conductance fluctuations (UCFs) and their amplitude reduction due to spin-orbit interaction, which is another important signature of phase coherent transport in strong spin-orbit interacting systems, in these 2D materials.

In this work, we report on a phase-coherent transport study of a $Bi_2O_2Se$ nanoplate and on observation of UCFs and their amplitude reduction due to spin-orbit interaction in this layered material. $Bi_2O_2Se$ nanoplates are grown via CVD on a mica substrate and our transport measurements are performed on a Hall-bar device, made from a CVD-grown $Bi_2O_2Se$ nanoplate on a $SiO_2/Si$ substrate, in a physical properties measurement system (PPMS). Here we note that the device is made from a $Bi_2O_2Se$ nanoplate with a thickness of 24 nm, slightly thicker than the nanoplates used previously in Ref. 13, in order to achieve a longer coherence length and thus stronger conductance fluctuations at low temperature measurements. In the following, we will first give brief descriptions about the materials growth and device fabrication. We then present the results of transport characterization measurements, in which the characteristic transport parameters, such as coherence length, spin relaxation length, and mean free path, and their temperature dependences in the nanoplates are extracted and discussed. In particular, we will show that a crossover between WAL and WL occurs at ~10 K in weak magnetic-field magnetotransport measurements of the nanoplate. After these characterization measurements, we present the main results of this work, namely the observation of UCFs at low temperatures. Here we will show for the first time that UCFs are observed over a large range of magnetic fields in a $Bi_2O_2Se$ nanoplate at low temperatures and that the amplitude of UCFs is reduced at temperatures below ~10 K when compared to the trend deduced from high temperature region. Our analysis shows that the observed amplitude reduction is fully in agreement with the prediction by the theory of Refs. 34 and 35, and can be firmly attributed to spin-orbit interaction effect. Our observation of UCFs and their amplitude reduction in the $Bi_2O_2Se$ nanoplate would stimulate further experimental and theoretical studies of the system.

    The $Bi_2O_2Se$ nanoplate studied in this work is grown in a low-pressure CVD system.[5] The source materials are $Bi_2O_3$ and $Bi_2Se_3$ (Alfa Aesar, 99.995%) powders, which are placed in a horizontal quartz tube. Fluorophlogopite mica is used as a growth substrate. For further details about the nanoplate growth, we would like to refer to Refs. 5 and 7. An optical image of a few CVD-grown $Bi_2O_2Se$ nanoplates on the mica substrate is shown in Fig. 1a. Here, the nanoplates seen with different brightnesses are different in thickness. The square and rectangular shapes of the nanoplates result from the $Bi_2O_2Se$ tetragonal crystal structure.[20] Figure 1b is an atomic force microscope (AFM) image of a $Bi_2O_2Se$ nanoplate with ultra-smooth surface. A high resolution transmission electron microscope (TEM) image of a $Bi_2O_2Se$ nanoplate is shown in Fig. 1c, confirming its high crystal quality. Figure 1d shows the typical spectra of X-ray photoelectron spectroscopy (XPS) measurements of as-grown

$Bi_2O_2Se$ nanoplates, consisting of characteristic Se 3d and Bi 4f peaks. The two peaks of Bi 4f located at 158.7 and 164.0 eV (higher than those in $Bi_2Se_3$) are clearly observed and are attributed to the chemical bonding of Bi(III)-$O_x$ in $Bi_2O_2Se$.[5, 21, 22] These results reveal the feature of chemical bond states of $Bi_2O_2Se$ nanoplates and indicate again the high quality of the prepared 2D crystals.

The as-grown $Bi_2O_2Se$ nanoplates are transferred from the growth mica substrate to a $SiO_2$/Si substrate for device fabrication using PMMA (polymethyl methacrylate)-mediated technique.[23] Devices are fabricated from selected $Bi_2O_2Se$ nanoplates on the $SiO_2$/Si substrate with contact electrodes made with a double layer of Ti/Au (5/70 nm in thickness) in a Hall-bar configuration using standard nanofabrication techniques of electron beam lithography, metal evaporation and lift-off, see Ref. 13 for further details about the device fabrication. The inset of Fig. 2a shows an optical image of the fabricated $Bi_2O_2Se$ nanoplate device studied in this work. A schematic of the transport measurement circuit setup is shown in the inset of Fig. 2b. The device is mounted to a rotatable sample holder and the measurements are performed in a PPMS equipped with a uniaxial magnet with supplied magnetic fields up to 9 T. The longitudinal voltage $V_x$ and the transverse voltage $V_y$ are detected simultaneously using standard lock-in technique with a 17 Hz, 100 nA excitation current $I$ applied between the S and D electrodes, see the inset of Fig. 2b. The sheet resistivity $R$ and the Hall resistance $R_{yx}$ are obtained as $R=R_{xx}\times W/L$ and $R_{yx}=V_y/I$, where $R_{xx}=V_x/I$ is the longitudinal resistance, $L$ is the distance between the two voltage probes along the current direction and $W$ is the width of the nanoplate. For the device studied in this work, we have $L$=4.2 µm and $W$=3.8 µm.

Figure 2a shows the measured sheet resistivity $R$ of the $Bi_2O_2Se$ nanoplate device shown in the inset of the figure as a function of temperature $T$ at zero magnetic field. It is seen that $R$ monotonically decreases as $T$ decreases from 300 to 10 K and saturates at $T$ below ~10 K. The Hall measurements show that the $Bi_2O_2Se$ nanoplate is electron conductive over the entire range of temperatures measured, which is consistent with the results previously reported.[13] Figure 2b shows the Hall electron mobility $\mu$ and sheet carrier density $n_{sheet}$ extracted from the measured $R$ and $R_{yx}$ at different temperatures (see Supplemental Materials). The sheet carrier density $n_{sheet}$ is seen to decrease with lowing $T$ from 300 K and tends to saturate when $T$ goes below ~10 K. Mobility $\mu$ increases as $T$ decreases and also shows a saturation behavior at $T$ below ~10 K. The temperature evolution characteristics of $\mu$ and $n_{sheet}$ suggest that phonon scattering to the cruising electrons is frozen out below ~10 K in the $Bi_2O_2Se$ nanoplate.[24] As shown in Fig. 2b, at 2 K, $n_{sheet}$ is $\sim 1.15 \times 10^{13}$ cm$^{-2}$ and $\mu$ reaches ~1466 cm$^2$/V·s, giving an electron mean free path of $l_e$ ~ 82 nm, about two orders of magnitude smaller than the length and the width of the Hall-bar device. Thus, the transport in the $Bi_2O_2Se$ nanoplate device is in the diffusive regime at 2 K.

Figure 3a shows the magnetoconductivity $\Delta\sigma_{xx}(B) = \sigma_{xx}(B) - \sigma_{xx}(B = 0)$ at 2 K with magnetic field $B$ applied at different tilting angles $\theta$, where $\theta$ is the angle between the field and the transverse in-plane direction as shown in the inset of Fig. 3b. Here, $\sigma_{xx}(B)$ is the longitudinal conductivity obtained from the measured longitudinal resistance $R_{xx}$ and transverse resistance $R_{yx}$ (see the Supplementary Materials for details), and the curves are successively vertically offset from that at $\theta = 90°$ for clarity. It is seen that when the magnetic field is applied perpendicular to the device substrate ($\theta = 90°$), the magnetoconductivity shows a sharp peak in the vicinity of zero field, i.e., a signature of WAL, consistent with the presence of strong spin-orbit interaction in the $Bi_2O_2Se$ nanoplate.[13, 25] When the magnetic field is rotated away from $\theta = 90°$ toward the transverse in-plane direction, the WAL peak becomes broadened. Figure 3b depicts the magnetoconductivity $\Delta\sigma_{xx}(B)$ against the vertical component of the field, i.e., $B\sin(\theta)$. Here, all the magnetoconductivity curves coincide with each other, verifying the 2D nature of transport in the $Bi_2O_2Se$ nanoplate. The magnetoconductivity measurements at low, perpendicular magnetic fields $B$ can be described by quantum interference theory,[26] which describes the quantum correction to the low-field magnetoconductivity in a diffusive 2D system as,[26, 27]

$$\Delta\sigma(B) = -\frac{e^2}{\pi h}\left[\frac{1}{2}\Psi\left(\frac{B_\varphi}{B}+\frac{1}{2}\right) + \Psi\left(\frac{B_{so}+B_e}{B}+\frac{1}{2}\right) - \frac{3}{2}\Psi\left(\frac{(4/3)B_{so}+B_\varphi}{B}+\frac{1}{2}\right) - \frac{1}{2}\ln\left(\frac{B_\varphi}{B}\right) - \ln\left(\frac{B_{so}+B_e}{B}\right) + \frac{3}{2}\ln\left(\frac{(4/3)B_{so}+B_\varphi}{B}\right)\right]. \quad (1)$$

Here, $\Psi(x)$ is the digamma function. $B_i$, with $i=\varphi$, SO, and $e$, describe the characteristic fields for different scattering processes and are given by $B_i = \hbar/(4eL_i^2)$ with $L_\varphi$ being the phase coherence length, $L_{SO}$ the spin relaxation length, and $L_e$ the electron mean free path.

Figure 3c shows the low-field magnetoconductivity curves of the $Bi_2O_2Se$ nanoplate device obtained at different temperatures. Here, the curves are successively vertically offset from that at $T = 2$ K for clarity. As the temperature increases from 2 to 12 K, the WAL peak gradually broadens and the magnetoconductivity curve eventually develops into a broad dip, a signature of WL. We fit the experimental data in Fig. 3c to Eq. (1) and the results are presented by solid lines in the figure. Figure 3d plots the transport characteristic lengths, $L_\varphi$, $L_{SO}$ and $L_e$, extracted from the fits against temperature. It is shown that $L_{SO}$ remains at a constant of ~250 nm, while $L_\varphi$ is ~500 nm at 2 K and decreases with increasing temperature as $L_\varphi \sim T^{-0.51}$ due to enhanced inelastic scattering at increased temperatures. This power-law temperature dependence is consistent with the 2D nature of transport in the nanoplate and proves that dephasing is dominantly due to electron-electron scattering processes with small energy transfers.[28]

The extracted values of $L_\varphi$ and $L_{SO}$ cross at $T$~10 K at which the magnetoconductivity shows a crossover between the WAL and WL characteristics. The extracted $L_{SO}$ is longer than the value (~150 nm) previously reported in Ref. 13 and thus the spin-orbit coupling strength of the device studied in this work is slightly weakened. This is because spin-orbit interaction is mainly derived from Bi-$p_z$ orbits and their contribution to the conduction electrons are weakened due to the fact that the $Bi_2O_2Se$ nanoplate (24 nm in thickness) used in this work is much thicker and the electron Fermi energy ($E_F$ ~19 meV) is significantly higher, when compared with the values (8 nm and 0.85 meV) in the nanoplate used in the previously reported work.[13] In addition, the extracted $L_\varphi$ and $L_e$ in the present device are also much longer than the corresponding values in the $Bi_2O_2Se$ nanoplate used in the previous work.[13]

Figure 4a shows the UCFs, $\delta G(B)$, extracted for the nanoplate device at different temperatures. Here, each curve is obtained by subtracting a polynomial background from the magnetoconductance data measured at vertically applied magnetic fields. When temperature increases, the main oscillation features in UCF patterns are reproducible, while the amplitude of the fluctuations gradually reduces due to enhanced inelastic scattering and thermal average. Fig. 4b shows the measured UCF patterns at different field orientations $\theta$ (cf. the inset of Fig. 3b). Here, the measured data are plotted against the vertical field component, i.e., Bsin(θ) and it is seen that the fluctuation patterns are solely dependent on the vertical component of the field, demonstrating again the 2D nature of transport in the Bi$_2$O$_2$Se nanoplate.[29] UCFs are a quantum interference effect and the phase coherence length $L_\varphi$ could also be extracted from the autocorrelation function of UCFs, $F(\Delta B) = \langle \delta G(B + \Delta B) \delta G(B) \rangle$, where the brackets $\langle \cdots \rangle$ represents an average over magnetic fields $B$.[30] For a 2D system, the phase coherence length $L_\varphi$ is estimated from $(L_\varphi)^2 \Delta B_c \sim \Phi_0$, where $\Phi_0$ is the flux quantum and $\Delta B_c$ is the width at half maximum of the autocorrelation function $F(\Delta B_c) = \frac{1}{2} F(0)$, see Supplementary Materials for further details. Figure 4c shows the extracted $L_\varphi$ at different temperatures. Here, again, $L_\varphi$ decreases with increasing temperature and can be fitted by a power law function of $L_\varphi \sim T^{-0.46}$. This temperature dependence is close to the result obtained based on WL and WAL analyses as shown in Fig. 3d and is consistent with the 2D nature of transport in the nanoplate, indicating that the same mechanism is involved in dephasing. However, we should note that the values of $L_\varphi$ extracted from UCFs are overall smaller than those extracted from the WL and WAL analyses. This could be understood as that unlike WL and WAL analyses, which dominantly take into account interference processes between specific, time reversal symmetric, closed paths, UCFs arise from interferences from all possible paths between two points in the Bi$_2$O$_2$Se nanoplates.[25, 30]

In a 2D system with $L > L_\varphi$ (as in the case for this work), the average effect among different coherent areas leads to a reduction in amplitude of conductance fluctuations. Considering this large-size average effect and the thermal average effect arising from finite temperatures, the fluctuation amplitude $\delta G_{rms}$, where subscript *rms* stands for the root-of-mean-square value, can be described by $\propto L_T/L \sqrt{\ln(L_\varphi/L_T)}$, where $L_T = \sqrt{\hbar D/k_B T}$ is the thermal length, without taking the amplitude reduction due to spin-orbit interaction into account.[31] Here, for a 2D system, $\delta G_{rms} \propto T^{-1/2}$ would be observed. However, as we will show in the following, spin-orbit interaction leads to a strong reduction in $\delta G_{rms}$ in our nanoplate device. Figure 4d shows $\delta G_{rms}/\langle G \rangle$ as a function of $1/\sqrt{T}$. The $\delta G_{rms}/\langle G \rangle$ exhibits a linear dependence on $1/\sqrt{T}$ in the high temperature region, where $L_\varphi$ is shorter than $L_{SO}$ and thus the effects of spin-orbit interaction on coherence transport phenomena are negligibly small. This $1/\sqrt{T}$ dependence is in good agreement with theory of electron-electron scattering induced dephasing in a 2D system.[30] The red dashed line in the figure is the result of extending the linear fit from the high temperature region to the low temperature region. It is clearly seen that at T < 10 K, the measured $\delta G_{rms}/\langle G \rangle$ increases more slowly and deviates away from the $1/\sqrt{T}$ dependence. This phenomenon has been previously observed in GaAs/AlGaAs heterojunctions[32] and InAs nanowires,[33] and has been attributed to the presence of strong spin-orbit interaction.[32, 33] In the Bi$_2$O$_2$Se nanoplate studied in this work, $L_{SO}$ becomes shorter than $L_\varphi$ at $T<$ 10 K, and strong spin-orbit coupling can thus influence the cruising electrons via introducing an effective magnetic flux which has opposite sign for spin-up and spin-down electrons, resulting in a reduction in amplitude of UCFs.[34, 35] To evaluate the effect of spin-orbital interaction on UCF amplitudes in our nanoplate, we fit our measured data in Fig. 4d with a semi-classical result,[32,34] $\delta G_{rms}/\delta G_{rms}^0 = 0.5\{3\exp(-4L_c^2/3L_{SO}^2) + 1\}^{1/2}$, where $\delta G_{rms}^0$ denotes the amplitude of UCFs in the system that would have no spin-orbital scattering (represented by the red dashed line in Fig. 4d for our device), $L_c$ is a characteristic length which is proportional to $L_T$. The blue solid line is the fitting result and $L_c = 2.8L_T$ is extracted from the fitting. As is shows in Fig. 4d, $\delta G_{rms}/\langle G \rangle$ at 2 K is reduced to ~75% of the value that would be obtained from the dashed line, i.e., if no spin-orbit coupling is present in the system. Note that in the infinitely strong spin-orbit coupling limit, a reduction factor of 50% in $\delta G_{rms}/\delta G_{rms}^0$ is predicted by the semi-classical theory.[34] Our observed reduction factor is about half of the value that would be observed in the infinitely strong spin-orbit interaction limit and is significantly larger than the reduction factor value observed previously in conventional GaAs/AlGaAs heterostructures.[32] Finally, we note that in an experiment, bias current heating could also induce a reduction in amplitude of UCFs. However, in the present work, it has been checked that the excitation current is small enough, so that such heating effect is negligible at temperatures considered in this work, see the Supplementary Materials for further details and discussion.

In conclusion, we have studied the phase-coherent transport in a Bi$_2$O$_2$Se nanoplate device. The device is made from a Bi$_2$O$_2$Se nanoplate of 24 nm in thickness, thicker than the nanoplate used in the study reported previously in Ref. 13. WAL and UCFs are observed in the magnetotransport measurements and the latter are for the first time reported for a 2D layered material. The measurements are well described by theory of coherent transport in 2D diffusive systems and characteristic transport lengths, $L_\varphi$, $L_{SO}$ and $L_e$, in the $Bi_2O_2Se$ nanoplate are extracted at different temperatures. It is found that at low temperatures of 2 to 12 K, $L_{SO}$ ~250 nm and $L_e$ ~82 nm are insensitive to temperature variations, while $L_\varphi$ shows a $T^{-1/2}$ temperature dependence suggesting that dephasing in the nanoplate arises dominantly from electron-electron scattering processes with small energy transfers. Most importantly, we find that the observed UCFs in the nanoplate show a strong reduction in fluctuation amplitude, which is in good agreement with a semi-classical description by taking into account strong spin-orbit scattering in the $Bi_2O_2Se$ nanoplate. It is expected that our results presented in this work will stimulate exploration of semiconductor $Bi_2O_2Se$ layers for applications in quantum electronics, spintronics and quantum information technology.

This work is supported by the Ministry of Science and Technology of China through the National Key Research and Development Program of China (Grant Nos. 2017YFA0303304, 2016YFA0300601, 2017YFA0204901, and 2016YFA0300802), and the National Natural Science Foundation of China (Grant Nos. 11874071, 91221202, 91421303, and 11274021). HQX also acknowledges financial support from the Swedish Research Council (VR).

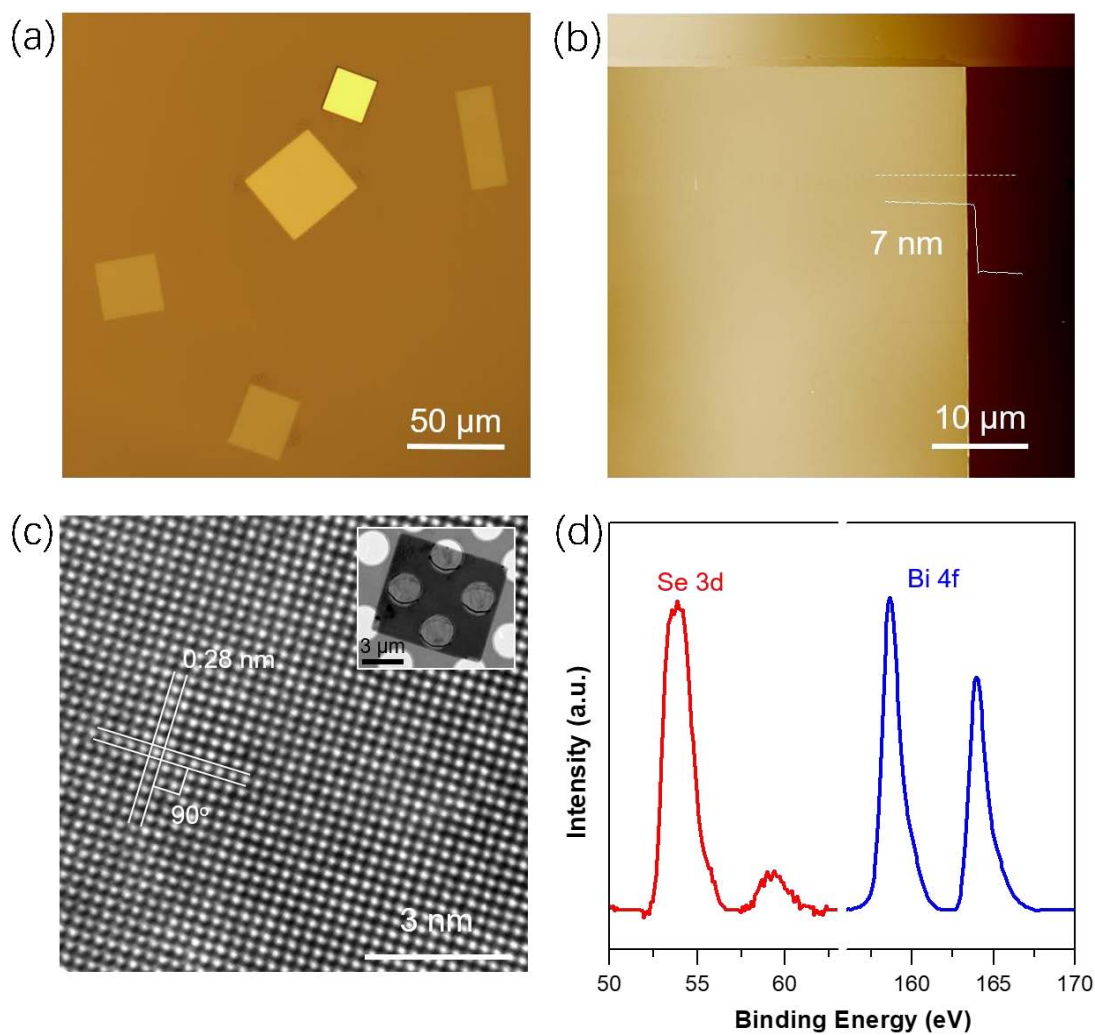

**Fig. 1** (a) Optical image of as-grown $Bi_2O_2Se$ nanoplates on a mica substrate. Here, a few of square- or rectangle-shaped nanoplates with different thickness, as indicated by their brightnesses, are seen. (b) Typical AFM image of a $Bi_2O_2Se$ nanplate, showing an ultra-smooth surface. (c) High resolution TEM image of a $Bi_2O_2Se$ nanoplate taken from the c-axis direction. The inset is a low-magnification TEM image of a $Bi_2O_2Se$ nanoplate transferred onto a holey carbon grid. (d) XPS characterization spectra of an as-grown $Bi_2O_2Se$ nanoplate sample, in which Bi 4f and Se 3d core level photoemission spectra are displayed.

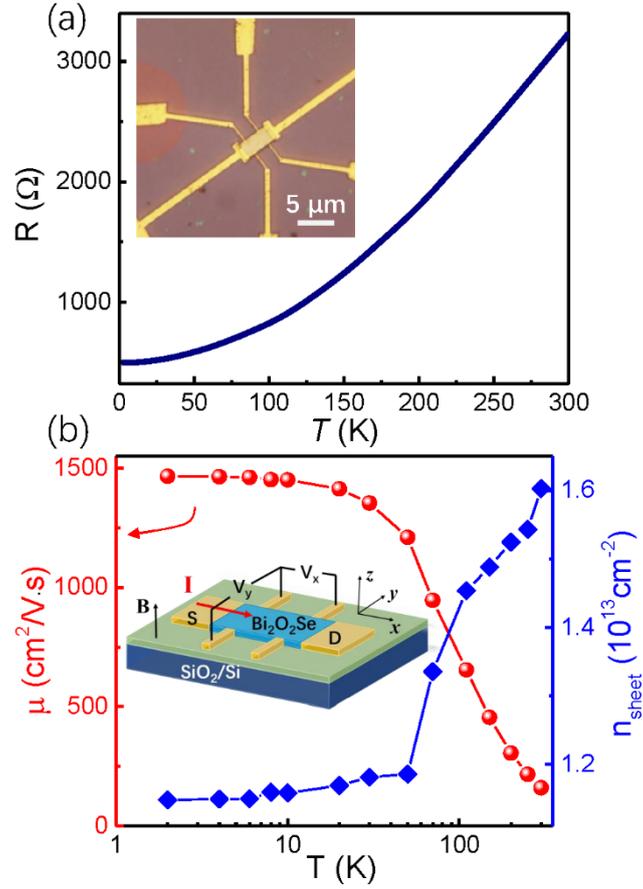

**Fig. 2** (a) Measured sheet resistivity $R$ of the $Bi_2O_2Se$ nanoplate studied in this work as a function of temperature $T$. The inset shows an optical image of the measured Hall-bar device made from the nanoplate. The nanoplate has a thickness of 24 nm. (b) Electron mobility ($\mu$) and sheet carrier density ($n_{sheet}$) extracted for the nanoplate as a function of temperature $T$. The inset shows a structure schematic of the $Bi_2O_2Se$ nanoplate device and the measurement circuit setup.

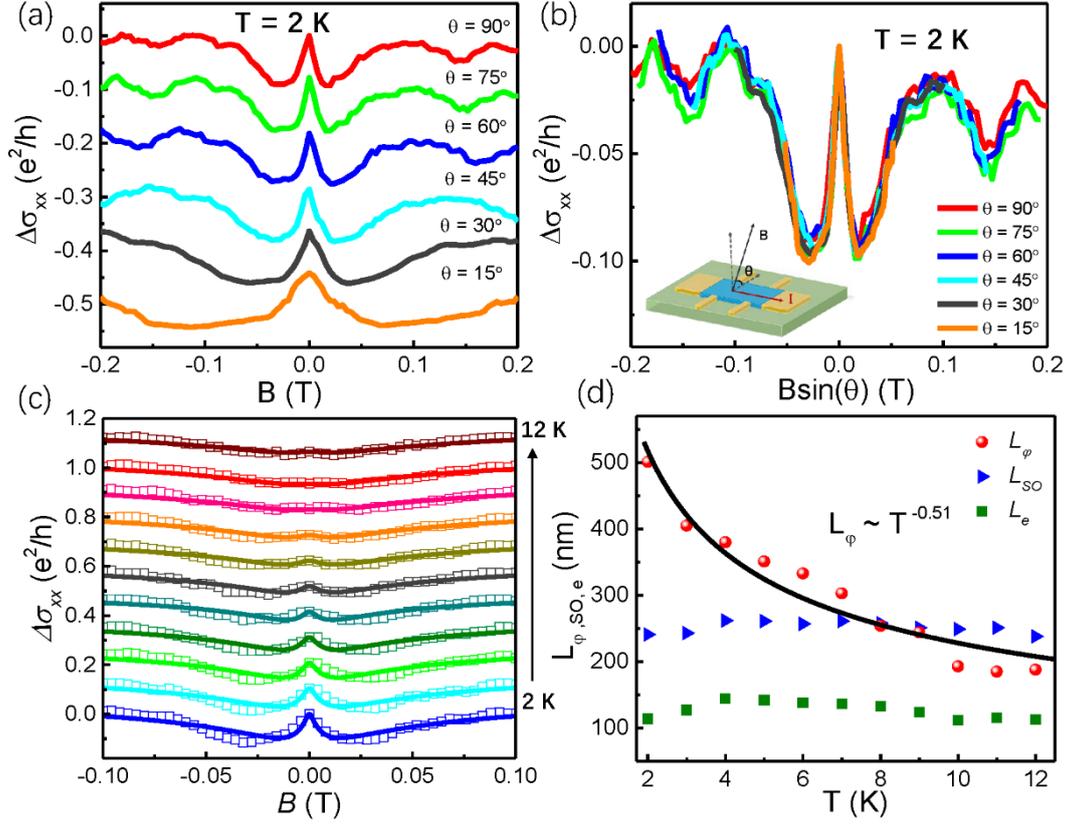

**Fig. 3** (a) Magnetoconductivity $\Delta\sigma_{xx}(B)$ measured at 2 K with the external magnetic field $B$ applied at different orientations $\theta$, see the schematic shown in the inset of (b) for the definition of $\theta$. The red curve displays the measurements at $\theta = 90°$ and all other measurements are successively vertically offset for clarity. (b) Magnetoconductivity $\Delta\sigma_{xx}(B)$ plotted against the vertical component Bsin(θ) of the applied magnetic field $B$. (c) Magnetoconductivity $\Delta\sigma_{xx}(B)$ measured for the $Bi_2O_2Se$ nanoplate device with the magnetic field applied in the vertical direction ($\theta = 90°$) at temperatures of 2 to 12 K. The bottom curve displays the measurement data at 2 K and all other curves are successively vertically offset for clarity. The solid lines are the results of theoretical fits of the experimental data to Eq. (1). Here, a WAL-WL crossover occurs as temperature increases. (d) Phase coherence length $L_\varphi$, spin relaxation length $L_{SO}$, and mean free path $L_e$, extracted from the fittings of the experimental results shown in (c), plotted against temperature $T$. The black solid line is a power-law fit of $L_\varphi$, showing $L_\varphi \sim T^{-0.51}$.

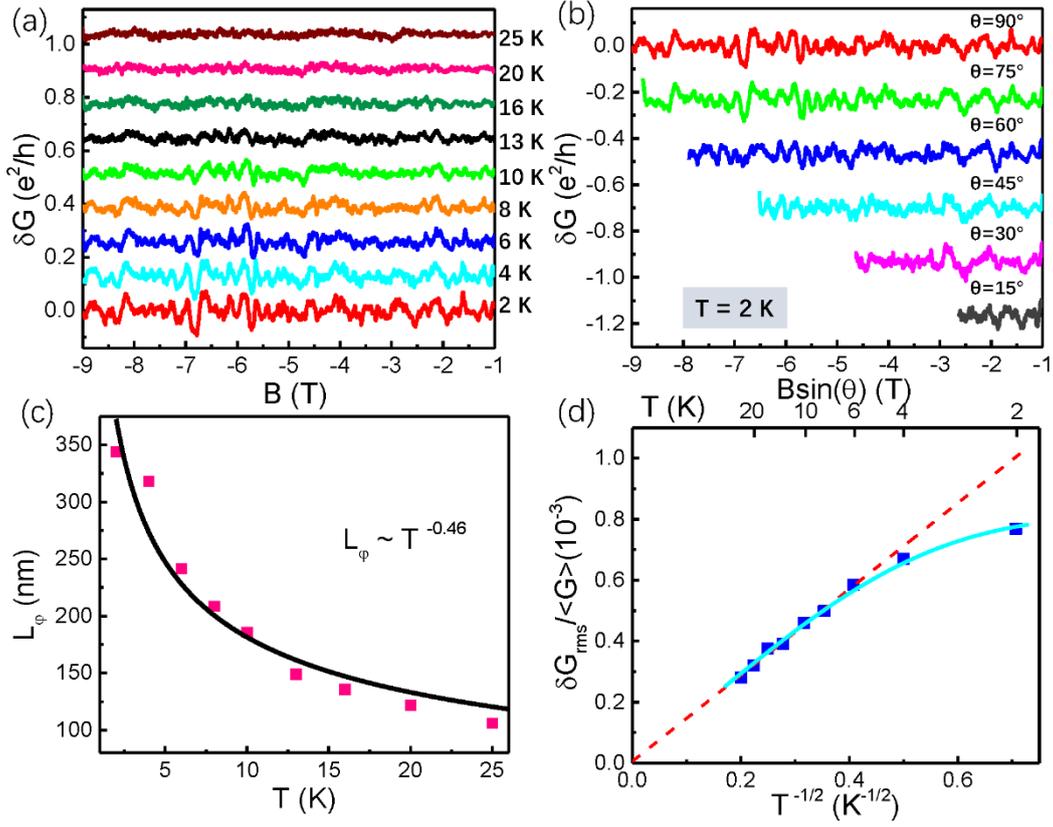

**Fig. 4** (a) Conductance fluctuations $\delta G$ plotted against vertically applied magnetic field $B$ at different temperatures. Similar aperiodic patterns appear in $\delta G$ at different temperatures. (b) Conductance fluctuations $\delta G$ measured with magnetic field applied in different orientations $\theta$ [cf. the inset of Fig. 3 (b)] as a function of the vertical component of the magnetic field. It can be seen that the conductance fluctuation patterns solely depend on the vertical component of the field. (c) Phase coherence length $L_\varphi$ extracted from the autocorrelation function $F(\Delta B)$ of the conductance fluctuations at different temperatures. The solid line is the result of power law fitting, showing that $L_\varphi \sim T^{-0.46}$. (d) Normalized root-of-mean-square value of the conductance fluctuations, $\delta G_{rms}/\langle G \rangle$, plotted against $T^{-1/2}$. The red dashed line represents the result obtained by a straight line fit to the trend found at high temperatures. The blue solid line is the result of theoretical fit to the measured data as discussed in the text. Deviations of the measured data from the red dashed line at low temperatures arise from reductions in UCF amplitude induced by spin-orbit coupling in the nanoplate.

# Supplementary Materials for

# Universal conductance fluctuations and phase-coherent transport in a semiconductor $Bi_2O_2Se$ nanoplate with strong spin-orbit interaction

Mengmeng Meng,[1] Shaoyun Huang,[1,*] Congwei Tan,[2,3] Jinxiong Wu,[2] Xiaobo Li,[1,3] Hailin Peng[2,*] and H. Q. Xu[1,4,5,*]

[1]*Beijing Key Laboratory of Quantum Devices, Key Laboratory for the Physics and Chemistry of Nanodevices and Department of Electronics, Peking University, Beijing 100871, China*

[2]*Center for Nanochemistry, Beijing National Laboratory for Molecular Sciences (BNLMS), College of Chemistry and Molecular Engineering, Peking University, Beijing 100871, China.*

[3]*Academy for Advanced Interdisciplinary Studies, Peking University, Beijing 100871, China*

[4]*Beijing Academy of Quantum Information Sciences, West Bld. #3, No.10 Xibeiwang East Rd., Haidian District, Beijing 100193, China*

[5]*NanoLund and Division of Solid State Physics, Lund University, Box 118, S-221 00 Lund, Sweden*

[*] Corresponding authors: Professor H. Q. Xu (hqxu@pku.edu.cn), Dr. Shaoyun Huang (syhuang@pku.edu.cn) and Professor Hailin Peng (hlpeng@pku.edu.cn)

## Contents

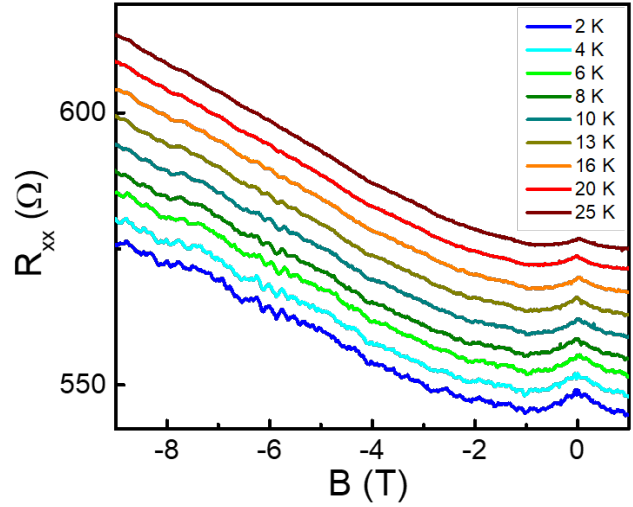

Figure S1. Measured longitudinal resistance ($R_{xx}$) of the same device as in the main article as a function of vertically applied magnetic field at temperatures of 2 to 25 K. For clarity, all the curves except for the one at 2 K are successively vertically offset. Here, the longitudinal resistance is obtained from $R_{xx} = V_x/I$ using lock-in technique [cf. the measurement circuit setup shown in the inset of Fig. 2(b) in the main article]. As we sweep the field, similar conductance fluctuations superposed on top of the background magnetoresistance are observable at different temperatures. The zoom-in plots in the vicinity of zero field show WAL characteristics at low temperatures and a WAL-WL crossover as temperature increases (cf. the results shown in the main article).

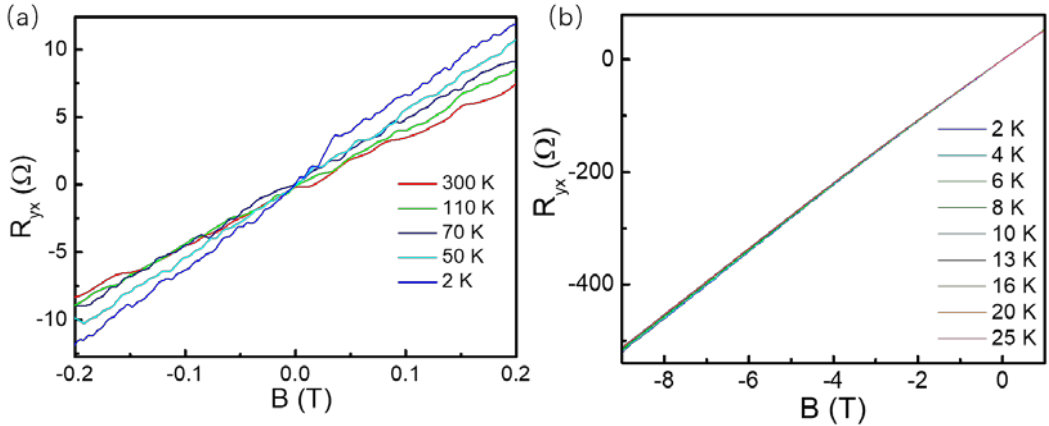

Figure S2. Transverse resistance ($R_{yx}$) of the device vs. vertically applied magnetic field $B$ at different temperatures. As the schematic shown in the inset of Fig. 2(b) in the main article, we measure $V_x$ and $V_y$ simultaneously using lock-in technique and the Hall resistance is obtained by $R_{yx} = V_y/I$. (a) $R_{yx}$ in a small, low-field region and (b) $R_{yx}$ over a large region of fields. The sheet carrier density $n_{sheet}$ is determined from the low-field Hall coefficient as $R_H = 1/ne$ with the Hall coefficient determined by the slope of $R_H = R_{yx}/B$.

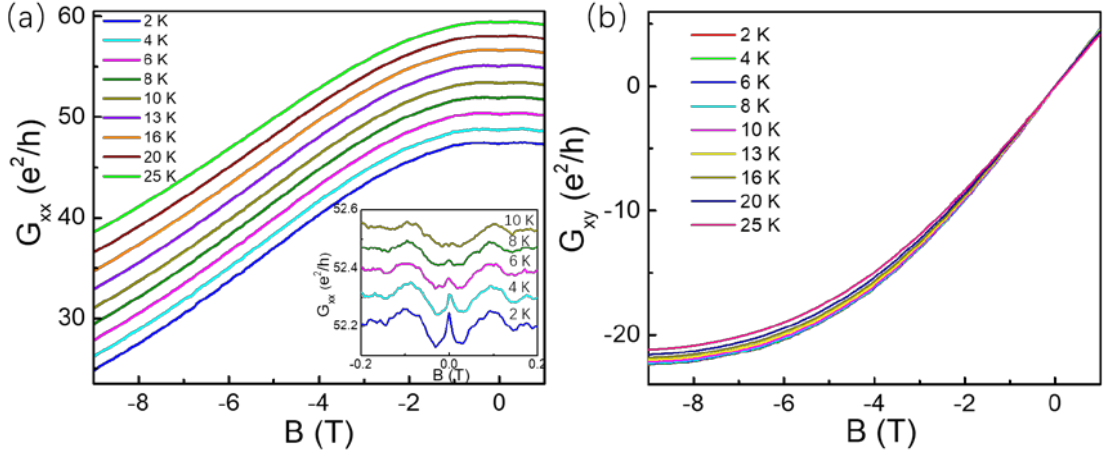

Figure S3. (a) Longitudinal conductance $G_{xx}$ of the device vs. vertically applied magnetic field $B$ at different temperatures. (b) Transverse conductance $G_{xy}$ of the device vs. vertically applied magnetic field $B$ at different temperatures. Here $G_{xx}$ and $G_{xy}$ can be calculated from measured $R_{xx}$ and $R_{yx}$ as $G_{xx} = \frac{R_{xx}}{R_{xx}^2 + R_{yx}^2}$ and $G_{xy} = \frac{R_{yx}}{R_{xx}^2 + R_{yx}^2}$. Longitudinal conductance fluctuations $\delta G$ shown in Fig. 4(a) in the main article are obtained by subtracting the polynomial backgrounds from the measured dada shown in (a). The inset in (a) shows zoom-in plots of the conductance around zero field. For clarity, all the curves in (a) except for the one at 2 K are successively vertically offset.

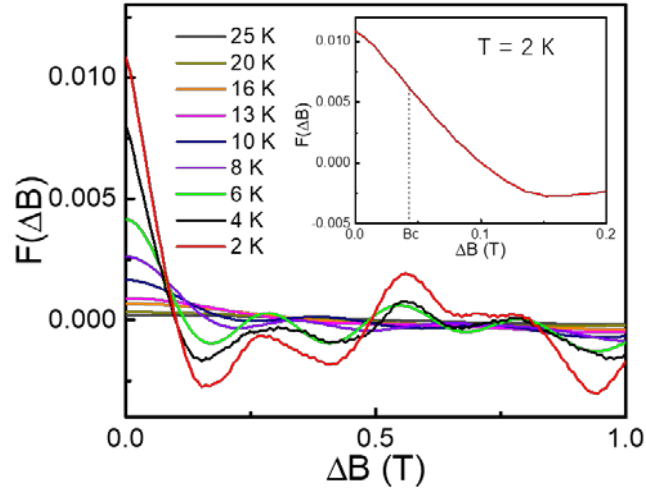

Figure S4. Autocorrelation function F($\Delta B$) of the measured conductance fluctuations shown in Fig. 4(a) in the main article. The characteristic field $B_c$ obtained at the half maximum $F(B_c) = \frac{1}{2} F(0)$ in each plot is related to phase coherence length $L_\varphi$ by $\Phi_0 = L_\varphi^2 B_c$, where $\Phi_0$ is the flux quantum. The inset shows a zoom-in plot of F($\Delta B$) at 2K, demonstrating how $B_c$ is determined from F($\Delta B$). The extracted values of $L_\varphi$ from the autocorrelation functions at different temperatures are shown in Fig. 4(c) in the main article.

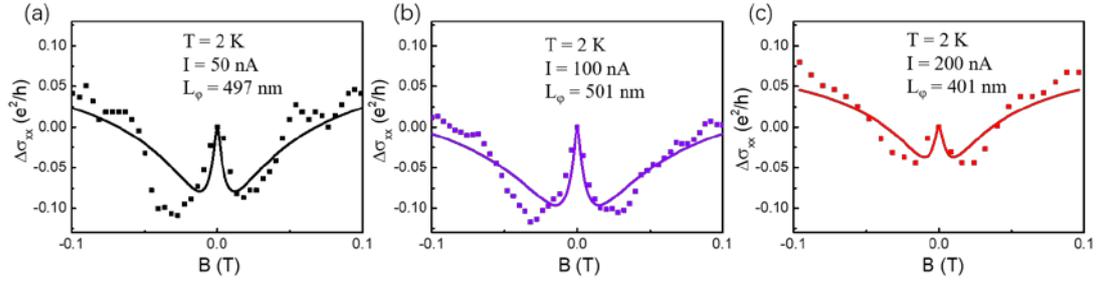

Figure S5. Magnetoconductivity of the same device as in the main article measured at 2 K with an excited current of (a) 50 nA, (b) 100 nA (the same as in Fig. 3c of the main article), and (c) 200 nA. Dots are measurement data and solid lines are the results of best fits of the measurement data to Eq. (1) of the main article. It is found that the extracted phase coherence length $L_\varphi$ of carriers from the measurements with an excited current of 100 nA is almost the same as that extracted from the measurements with a 50 nA excitation current, but is much larger than the value extracted from the measurements with a 200 nA excitation current. Thus, the effect of bias current heating on the phase coherent length in the device is negligible in the temperature range considered in the main article.